\documentclass[aps,prl,twocolumn,superscriptaddress,showpacs,floatfix]{revtex4-1}
\usepackage{graphicx,bm,epsfig,textcomp,color,dcolumn,setspace,array,mathrsfs,amsmath,amssymb,gensymb,booktabs,url,hyperref,bibentry,natbib}
\begin{document}

\title{Spin fluctuations in quantized transport of magnetic topological insulators}

\author{Yu-Hang Li}
\email[]{yuhang.li@ucr.edu}
\affiliation{Department of Electrical and Computer Engineering, University of California, Riverside, California 92521, USA}
\author{Ran Cheng}
\email[]{rancheng@ucr.edu}
\affiliation{Department of Electrical and Computer Engineering, University of California, Riverside, California 92521, USA}
\affiliation{Department of physics, University of California, Riverside, California 92521, USA}

\begin{abstract}
In magnetic topological insulators, quantized electronic transport is interwined with spontaneous magnetic ordering, as magnetization controls band gaps, hence band topology, through the exchange interaction. We show that considering the exchange gaps at the mean-field level is inadequate to predict phase transitions between electronic states of distinct topology. Thermal spin fluctuations disturbing the magnetization can act as frozen disorders that strongly scatter electrons, reducing the onset temperature of quantized transport appreciably even in the absence of structural impurities. This effect, which has hitherto been overlooked, provides an alternative explanation of recent experiments on magnetic topological insulators.
\end{abstract}

\maketitle

The inquiry into topological materials has recently mingled with the quest for low-dimensional magnets, giving birth to an emerging frontier known as magnetic topological insulators (TIs) where a topologically non-trivial band gap is controllable by spontaneous magnetic ordering~\cite{Liu2016The,Tokura2019Mangetic,Qi2011Topological,Hasan2010Colloquium}. Therefore, manipulating magnetization becomes a new tuning nob of the quantized electronic transport. For example, in a TI with coexisting ferromagnetic order, the system should exhibit the quantum anomalous Hall (QAH) effect when a finite magnetization is established below the Curie temperature ($T_c$)~\cite{Yu2010Quantized}. However, the QAH effect was first realized in a magnetically doped TI in which the magnetic moments are embedded randomly~\cite{Chang2013Experimental}, leading to strong disorder effects that significantly reduce the electron mobility hence inhibit the appearance of quantized transport~\cite{Chang2015Zero,Checkelsky2012Dirac,Chang2013Thin,Kou2013Interplay}. As a result, the actual onset temperature of QAH effect in such a material is much lower than the magnetic ordering temperature.

Removing this road block calls for magnetic TIs in which the magnetic moments are arranged periodically on a lattice. This can be achieved in either an intrinsic magnetic TI~\cite{Zhang2019Topological,Deng2020Quantum,Liu2020Robust,Ge2020High} or a heterostructure with a TI sandwiched between two magnetic thin films~\cite{Qi2008Topological,Watanabe2019Quantum}. However, the quantized transports in these systems turned out to be as vulnerable to an increasing temperature as those studied in magnetic doped TIs~\cite{Mogi2015Magnetic}. While this discouraging observation might still be attributed to structural impurities, it remains an open question what is responsible for the disappearance of QAH effect at a temperature far below $T_c$.

In this Letter, we introduce an alternative mechanism in magnetic TIs that can substantially reduce the onset temperature of quantized transport even in the absence of structural impurities. Contrary to the electrons governed by an formidably high Fermi temperature, spin fluctuations disturbing the magnetic order are very susceptible to thermal agitations~\cite{nolting2009quantum}. Because spin fluctuations take place on a time scale that is orders of magnitude larger than the electron relaxation time~\cite{marder2010condensed}, the electron dynamics can adjust \textit{adiabatically} to the instantaneous configuration of magnetic moments, seeing the instantaneous spin fluctuations as a \textit{random potential} almost frozen in time. For this reason, thermal spin fluctuations in the magnetic degree of freedom can manifest as effective disorders affecting the electron transport, even though magnetic atoms are arranged perfectly on a lattice free of structural impurities.

\begin{figure}[t]
  \centering
  \includegraphics[width=\linewidth]{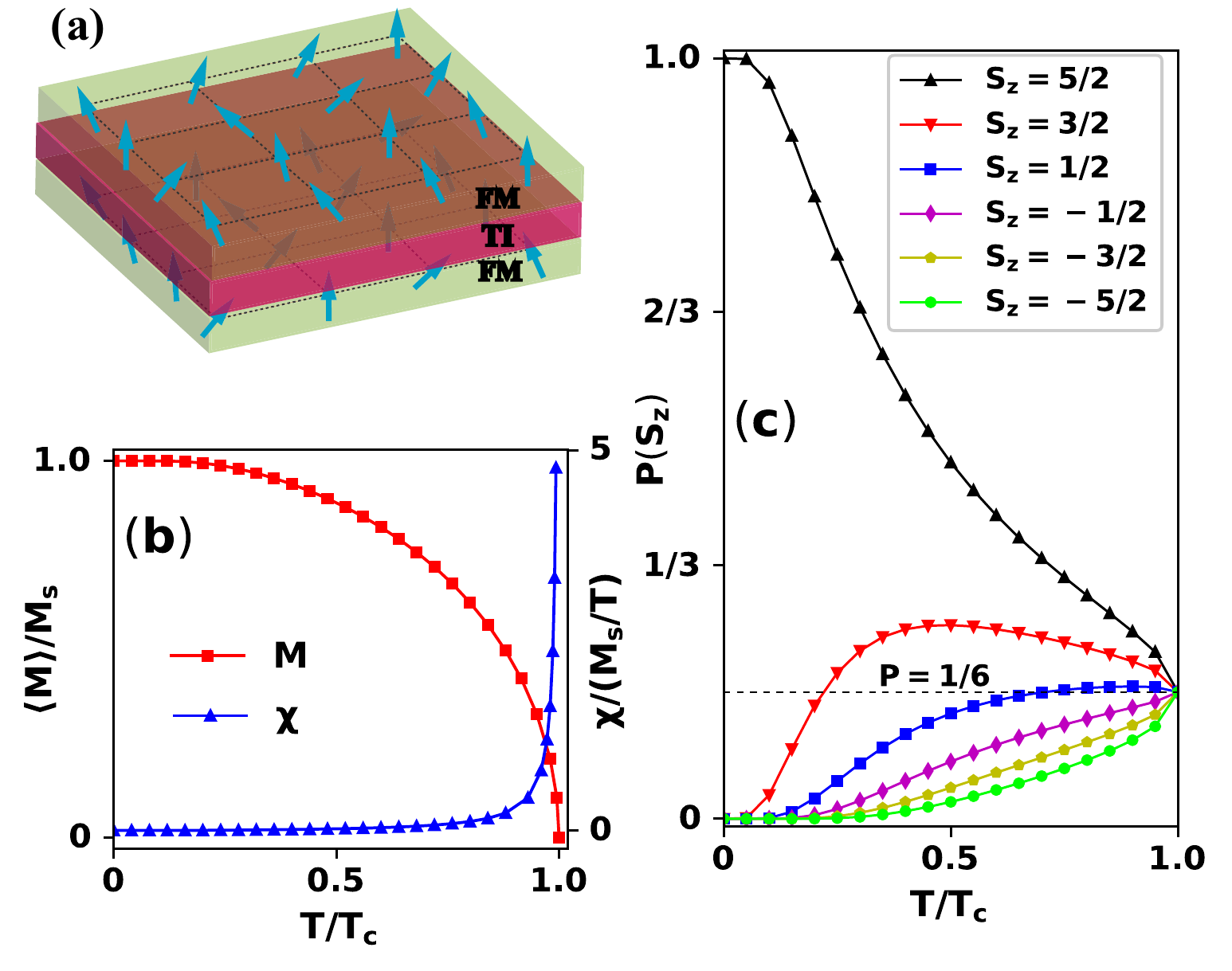}
  \caption{(a) Schematic of a magnetic TI in the presence of spin fluctuations.
  	(b) The mean field scaled by $M_s\equiv \langle M(T\rightarrow0)\rangle$ for $B\rightarrow0$ and the susceptibility $\chi$ as functions of temperature.
	(c) Probabilities of different $S_z$ on an individual spin versus temperature for $S=5/2$.
  	}
\label{Schemat_and_Mag}
\end{figure}

As schematically illustrated in Fig.~\ref{Schemat_and_Mag}(a), we model the system as a magnetic trilayer where topological electrons are confined between two magnets, which applies to not only a heterostructure but also an intrinsic magnetic TI with uniform magnetic ordering~\cite{Fu_comment}. To ensure the relative orientation of the two magnetic layers, we include an auxiliary magnetic field $B$ along $z$ axis to stabilize the system, but the $B\rightarrow0$ limit will be taken at the end. Now let us quantify the magnetization dressed with spin fluctuations in an individual magnetic layer, which is supposed to be independent of all other layers as schematically illustrated in Fig.~\ref{Schemat_and_Mag}. The minimal Hamiltonian of the magnet considered here is
\begin{align}
    \mathcal{H}_M=-J\sum_{\left<ij\right>}\bm{S}_i\cdot\bm{S}_j-\kappa\sum_i S_{i,z}^2-g\mu_B B\sum_iS_{i,z},
\end{align}
where $J>0$ is the (intralayer) Heisenberg exchange coupling, $\kappa$ is the uniaxial anisotropy, $g$ is the Land\'{e} factor, $\mu_B$ is the Bohr magneton, and $\langle ij\rangle$ enumerates all nearest-neighbors. The spin vector $\bm{S}_i$ is dimensionless.

In the mean-field approximation~\cite{nolting2009quantum}, spins become effectively decoupled while the exchange interaction that entangles different spins recasts as an effective mean field $\langle M\rangle=J\langle \sum_i S_{i,z}\rangle_T/(g\mu_B N)$ where $N$ is the total number of spins and $\langle\cdots\rangle_T$ denotes the thermal average. Consequently, the system becomes a paramagnet interacting with a total magnetic field $B_{\rm tot}=B+\langle M\rangle$ as if there is no exchange interaction. In the limit $J\gg\kappa$, the effective Zeeman energy is $E=-g\mu_{B}\left(B+\left<M\right>\right)\sum_iS_{iz}$, from which the mean field $\langle M\rangle$ can be solved self-consistently~\cite{nolting2009quantum}. Figure~\ref{Schemat_and_Mag}(b) shows the mean field $\langle M\rangle$ and the susceptibility $\chi\equiv\lim\limits_{B\to 0}{[\langle M(B)\rangle-\langle M(0)\rangle]/B}$ for $S=5/2$ as a function of temperature scaled by the Curie temperature $T_c=aJS\left(S+1\right)/3k_B$ on a simple square lattice with the coordination number $a=4$.  As every spin is now isolated from all other spins, the probability of an individual spin $\bm{S}_i$ taking $S_z$ perpendicular to the plane is determined straightforwardly by the Boltzmann distribution $P\left(S_z\right)=\exp{\left(-\varepsilon/k_{B}T\right)}/Z$ where $\varepsilon=-g\mu_BS_z(B+\langle M\rangle)$ and the partition function $Z=\sinh{\left[\left(2S+1\right)y\right]/\sinh{y}}$ with $y=aJ\left<M\right>/2T$. As plotted in Fig.~\ref{Schemat_and_Mag}(c), the spin is fully polarized to $S_z=S$ at $T=0$, whereas when $T \rightarrow T_c$ all possible quantized values of $S_z$ tend to be equally probable, destroying the magnetization completely at $T_c$.

The mean-field approach enables us to determine the projection of a given spin $\bm{S}_i$ on z-direction probabilistically. With the spherical parameterization $\bm{S}_i=S( \sin{\theta_i}\cos{\phi_i},\ \sin{\theta_i}\sin{\phi_i},\ \cos{\theta_i})$, it amounts to determining $\theta_i$ probabilistically. The azimuthal angle $\phi_i$, on the other hand, cannot be captured by the mean-field picture. Because we only consider the incoherent thermal spin fluctuations, $\phi_i$ should be uniformly distributed within the range $[0,2\pi)$. Moreover, because different modes of spin excitation superimpose with completely random phases, $\phi_i$ should be independent of its neighbors. In other words, the variable $\phi$ is spatially uncorrelated, or $\langle \phi_i(t)\phi_j(t)\rangle\sim\delta_{ij}$ at any instant of time. In contrast, the temporal correlation of $\phi$ is much larger than the electron relaxation time. Specifically, $\langle\phi_i(t)\phi_i(t')\rangle\sim e^{-|t-t'|/\tau_s}$, where the characteristic decay time $\tau_s$ may depend on the mode of excitation, but a qualitative estimation is that $\tau_s\sim1/\alpha\omega$ where $\alpha$ is the Gilbert damping and $\omega$ is the frequency of ferromagnetic resonance. So a typical value of $\tau_s$ is on the order of $10-100$ns. Comparatively, the electron relaxation time $\tau_e$ determined by the Fermi energy is on the order of $1-10$fs, which is $7$ orders of magnitude smaller than $\tau_s$. A similar argument applies to the correlation of $\theta$ as well. Therefore, while spin fluctuations are spatially uncorrelated, they exhibit extremely long temporal correlation, which amounts to a random potential frozen in time acting on the electrons~\cite{domain}. This justifies the adiabatic approximation essential to our following discussions.

Even though Dirac electrons and magnetic layers repeat periodically in an intrinsic magnetic TI, the system can be simplified as a trilayer heterostructure consisting of only one TI layer sandwiched between two magnetic layers as illustrated in Fig.~\ref{Schemat_and_Mag}(a)~\cite{Fu_comment}. Under the basis $\psi_{\bm{k}}=(c_{\bm{k}\uparrow}^{t},\ c_{\bm{k}\downarrow}^{t},\ c_{\bm{k}\uparrow}^{b},\ c_{\bm{k}\downarrow}^{b})^{T}$ with $c_{\bm{k}\sigma}^{t\left(b\right)}$ annihilating an electron of momentum $\bm{k}$ and spin $\sigma$ on the top (bottom) surface, the magnetic TI can be described by the Hamiltonian $\mathcal{H}_{MTI}=\mathcal{H}_{TI}+\mathcal{H}_{ex}$, where~\cite{Yu2010Quantized,Nomura2011Surface,Liu2009Magnetic}
\begin{align}
\mathcal{H}_{TI}&=v_{F}\left(k_{y}\tau_{z}\otimes\sigma_{x}-k_{x}\tau_{z}\otimes\sigma_{y}\right)+m(\bm{k})\tau_{x}, \label{eq:HTI}\\
\mathcal{H}_{ex}&=J_{ex}\sum\nolimits_i\bm{S}_i\cdot\bm{\sigma}. \label{eq:ex}
\end{align}
Here, $v_F$ is the Fermi velocity, $J_{ex}$ is the exchange coupling between the Dirac electrons and the magnetic moments, $m\left(\bm{k}\right)=m_{0}+m_{1}k^2$ describes the overlap of Dirac electrons in the top and bottom surfaces, and $\bm{\sigma}$ and $\bm{\tau}$ are the vectors of Pauli matrices acting on the spin and layer degree of freedom, respectively. The lattice wave vectors $k_{x,y}$ are defined in the first Brillouin zone of a $L\times W$ square lattice with the lattice constant $a\equiv1$. Since the Fermi temperature $T_F$ is orders of magnitude larger than $T_c$, the electron dynamics is effectively in the zero temperature regime as we focus on $T<T_c$~\cite{Thermal,Otrokov2019Prediction}. Unless otherwise stated, we will take $v_{F}=1$ as the energy unit and assume $m_{1}=1$, $k_BT_{c}=0.002$, $J_{ex}=0.035$ and $S=5/2$.

\begin{figure}[t]
  \centering
  \includegraphics[width=\linewidth]{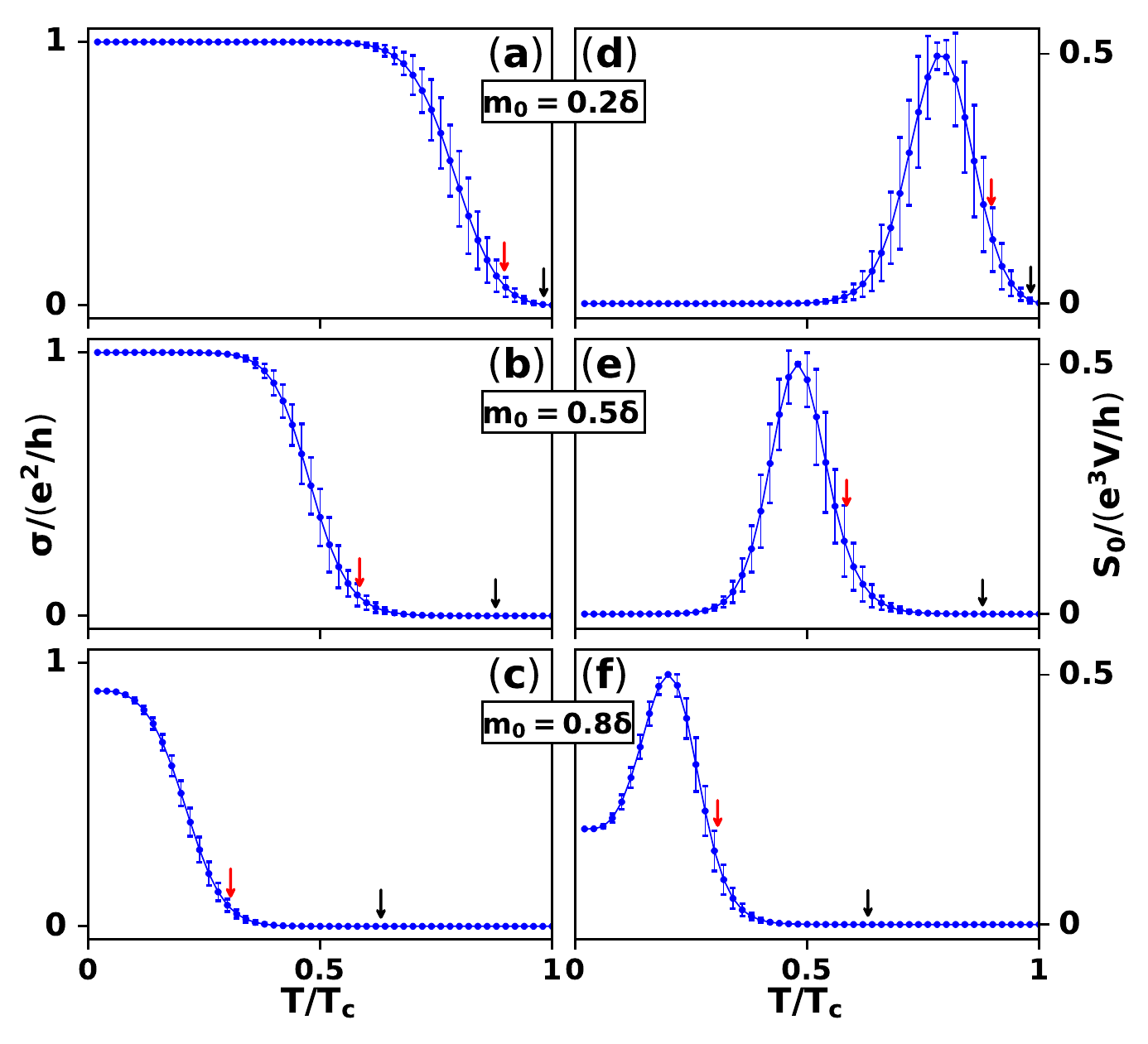}
  \caption{ 
  	(a)-(c): Ensemble average of the two-terminal conductance $\sigma$ as a function of temperature for different $m_{0}$ and fixed $\delta=g\mu_BJ_{ex}M_s$ (the maximum exchange field). (d)-(f): the corresponding zero-frequency current noise $S_0$. The red arrows mark the critical temperature $T_{sf}$ obtained by the finite-size scaling shown in Fig.~\ref{Hall_PDiagram}. The black arrows mark where $m_0=\lambda$, representing the critical temperature $T_{hm}$ in the absence of spin fluctuations. The system size is $L=W=200$ and the error bars are magnified ten times for visual clarity.
   	}
\label{cond_noise}
\end{figure}

To demonstrate the influence of spin fluctuations on the electron transport more clearly, it is instructive to first look into the homogeneous case without any spin fluctuations, in which $S_z$ is described by the mean field while $S_x$ and $S_y$ are completely ignored. In this situation, the lattice periodicity is restored in the exchange field, so we can transform the exchange Hamiltonian in Eq.~\eqref{eq:ex} into the momentum space, and $\mathcal{H}_{MTI}(\bm{k})=\mathcal{H}_{TI}+\lambda\tau_{0}\otimes\sigma_{z}$, where $\lambda=g\mu_BJ_{ex}\langle M\rangle$ is the homogeneous exchange field that depends on temperature through the mean field $\langle M\rangle$. Diagonalizing $\mathcal{H}_{MTI}(\bm{k})$ gives the band dispersion and the corresponding eigenstates, based on which we can calculate the Chern numbers characterizing different topological phases. At low temperatures, $\lambda>m_{0}$, the system is a QAH insulator with a Chern number $\mathcal{C}=1$. By contrast, the system becomes a normal insulator (NI) with $\mathcal{C}=0$ when $\lambda<m_0$ at high temperatures. Setting $\lambda=m_0$ solves the critical temperature $T_{hm}$ for the homogeneous case. Therefore, the system undergoes a topological phase transition at finite temperature below $T_c$ only if $m_0$ is less than the maximum exchange field $\delta\equiv g\mu_BJ_{ex}M_s$ with $M_s=\langle M(T\rightarrow0)\rangle$ the saturated mean field. In Fig.~\ref{cond_noise}, the critical temperature $T_{hm}$ for the homogeneous case is marked by the black arrows for different ratios of $m_0/\delta$.

Next, we turn to the transport property in the presence of spin fluctuations, which, as discussed above, act on electrons as a frozen random potential. In the considered magnetic TI, the appearance of topological edge states can be minimally revealed in a two-terminal junction, where the longitudinal conductance is $\mathcal{\sigma}=e^2/h$ ($\sigma=0$) in the QAH (NI) phase. We calculate $\sigma$ through the Landauer-B\"uttiker formula~\cite{Ryndyk2016Theory} $\sigma=\text{Tr}\left[\Gamma_{L}G^{r}\Gamma_{R}G^{a}\right]$, where $\Gamma_{\beta}=i\left[\Sigma_{\beta}^{r}-\left(\Sigma_{\beta}^{r}\right)^{\dagger}\right]$ with $\beta=L$ or $R$, and $G^{r}=\left(G^{a}\right)^{\dagger}=\left(E_{F}-H_{MTI}-\Sigma_{L}^{r}-\Sigma_{R}^{r}\right)^{-1}$ with $E_{F}$ the Fermi energy and $\Sigma_{\beta}^{r}$ the self energy due to the coupling with metallic leads. 

To simulate the random potential, we generate a set of $L\times W=200\times200$ random numbers representing $S_z=S\cos\theta$ on each lattice according to the probability distribution $P(S_z)=\exp(-\varepsilon/k_BT)/Z$ determined by the mean-field approach. We also assign each spin a random phase $\phi$ specifying its transverse component as discussed previously. Then we calculate the conductance  $\sigma$ under this particular configuration of random potential. Repeating this procedure for $160$ times, we obtain the ensemble average of  $\sigma$, which is shown in Fig.~\ref{cond_noise}(a)-(c) as a function of temperature for different $m_0$. We see that  $\sigma$ changes from $e^2/h$ to $0$ (\textit{i.e.}, transition from the QAH to NI phase) at a critical temperature $T_{sf}$ manifestly below what it would be without spin fluctuations (\textit{i.e.}, $T_{hm}$ determined by solving $\lambda=m_0$), as indicated by the red arrows. The reduction of critical temperature appears to be more striking for larger $m_0$ in Fig.~\ref{cond_noise}. For $m_0=0.8\delta$ [Fig.~\ref{cond_noise}(c)], $\sigma$ even becomes ill quantized in the QAH phase due to the finite-size effect~\cite{EdgeWidth}. If the system is infinite, $\sigma$ would be a step function across the critical point. Finite-size effects will be discussed in more detail later.

The topological phase transition between the QAH insulator and the NI can be alternately characterized by the current noise $S\left(\omega\right)=\frac{1}{2}\int{d\tau}e^{i\omega\tau}\langle\delta\hat{I}\left(t\right)\delta\hat{I}\left(t+\tau\right)+\delta\hat{I}\left(t+\tau\right)\delta\hat{I}\left(t\right)\rangle$, where $\delta\hat{I}\left(t\right)=\hat{I}\left(t\right)-\langle\hat{I}\left(t\right)\rangle$ with $\hat{I}\left(t\right)$ the current operator~\cite{Blanter2000Shot,Martin2005Course}. Using the non-equilibrium Green's function~\cite{Li2018Noise}, we calculate the zero-frequency current noise $S_0$. Figure.~\ref{cond_noise}(d)-(f) show the ensemble average of $S_0$ corresponding to Figs.~\ref{cond_noise}(a)-(c). The noise $S_0$ peaks at the critical point and extends over a finite range of temperature due to finite-size effects; it will become infinitely sharp at the critical point if the system is infinite. We see that $\sigma$ and $S_0$ plotted in Fig.~\ref{cond_noise} perfectly agree with the relation $S_0=2e^3V\sigma\left(1-\sigma\right)/h$ where $V$ is the bias voltage across the junction, affirming that the QAH edge states can be described by a one-channel ballistic tunneling model~\cite{Martin2005Course}.

Without spin fluctuations, the mean field $\langle M\rangle$, hence the exchange field $\lambda$, decreases as temperature is raised. When $\lambda$ becomes comparable to $m_0$, the chiral edge states on opposite transverse edges start to overlap, merging into the bulk states~\cite{EdgeWidth}. This destroys the electron transport and diminishes the conductivity. Spin fluctuations as random potential, on the other hand, brings about scattering of the chiral edge states, which facilitates their overlapping and merging into the bulk states, so the phase transition takes place at a reduced temperature. This subtle mechanism can be unraveled by studying the non-equilibrium current distribution inside the magnetic TI. Under a bias voltage $V$ across the system, the local current flowing from site $i$ to its neighbor $j$ is given by $\bm{J}_{i\rightarrow j}^{ne}=\text{Im}\left\{\text{Tr}\left[\hat{t}_{ij}\left(G^{r}\Gamma_{L}G^{a}\right)_{ji}\right]\right\}2e^2V/h$ where $\hat{t}_{ij}$ is the hoping matrix~\cite{Jiang2009Numerical}.

\begin{figure}[t]
  \centering
  \includegraphics[width=0.5\textwidth]{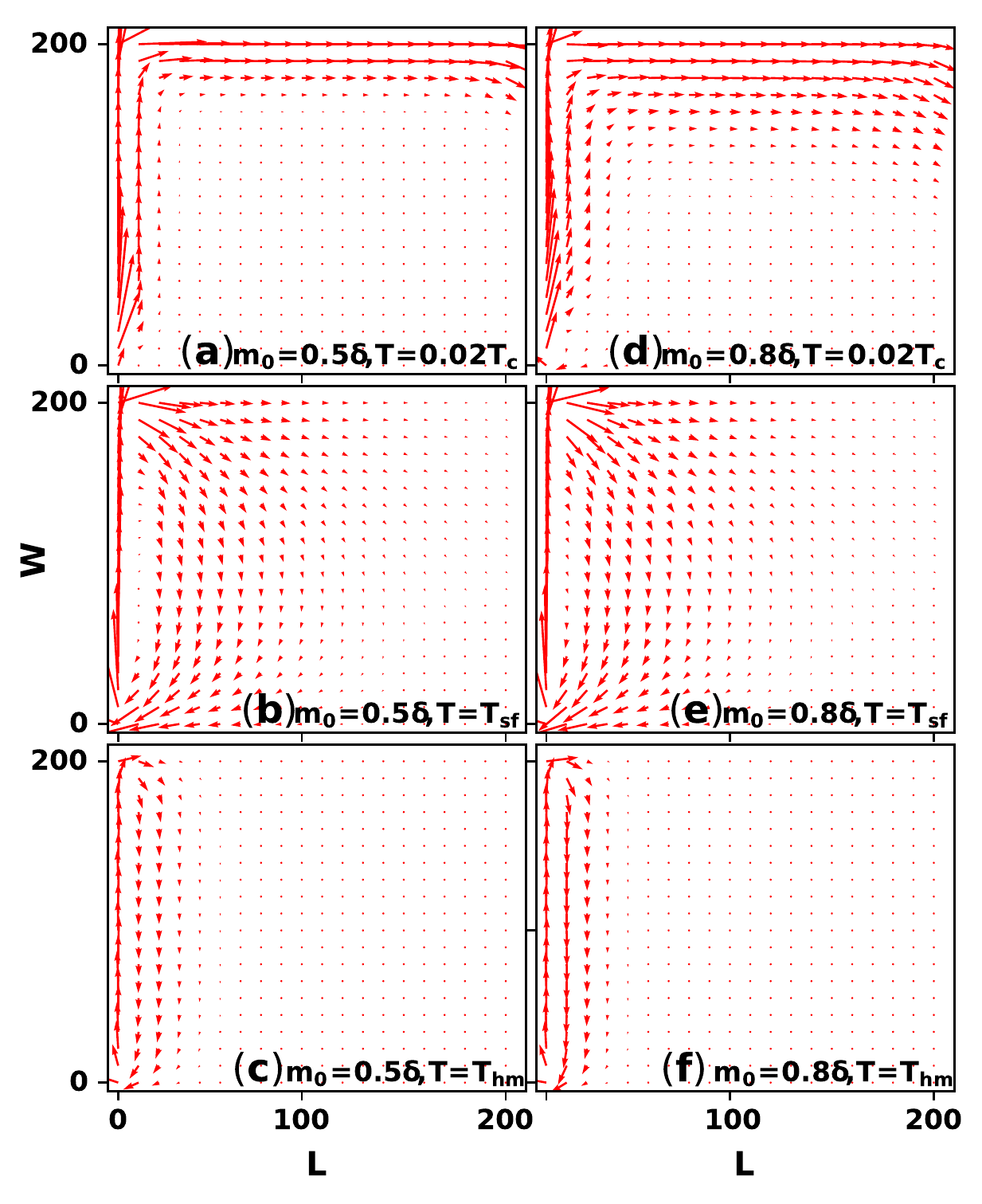}
  \caption{Non-equilibrium current distributions for $m_0=0.5\delta$ (a)--(c) and $m_0=0.8\delta$ (d)--(f) at $T=0.02T_c$, $T=T_{tsf}$, and $T=T_{hm}$. Red arrows indicate local current densities and directions.}
\label{dist}
\end{figure}

Figure~\ref{dist} shows the distributions of non-equilibrium currents in the TI at three representative temperatures for $m_0=0.5\delta$ [(a)--(c)] and $m_0=0.8\delta$ [(d)--(f)], respectively. At $T\ll T_{sf}$ and $m=0.5\delta$ [Fig.~\ref{dist}(a)], the electron flow is fully confined to one edge, so the conductance is quantized--a hallmark of the QAH effect. For $m=0.8\delta$[Fig.~\ref{dist}(d)], however, the edge current becomes much wider so that it partially leaks into the opposite edge and flows backwards, leading to an ill-quantized conductance as shown in Fig.~\ref{cond_noise}(c). At the true critical point $T=T_{sf}$ [(b) and (e)] where $\lambda>m_0$, spin fluctuations strongly scatter the electrons from one edge to the other, because of which electrons cannot propagate in one direction dictated by the applied bias voltage; they are instead back-scattered to the left lead. Accordingly, the chiral edge states become indistinguishable from the bulk states. At $T=T_{hm}$ [(c) and (f)] where $\lambda=m_0$, the edge states completely disappear and the conductance is identically zero. Integrating the current density over the full width $W$ yields a conductance that quantitatively agrees with the results shown in Fig.~\ref{cond_noise}, confirming the validity of the non-equilibrium distribution.

\begin{figure}[t]
  \centering
  \includegraphics[width=\linewidth]{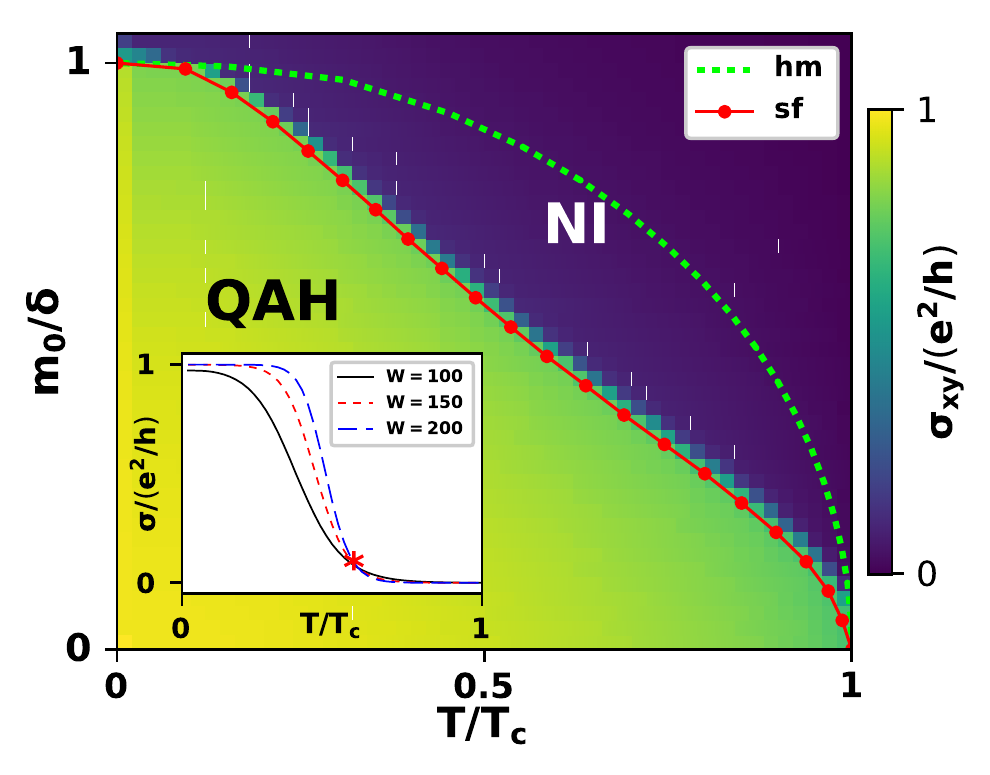}
  \caption{Phase diagram of the two-terminal conductance on the $m_0-T$ plane. The inset illustrates how $T_{sf}$ is obtained from finite-size scaling. The red dots plot $T_{sf}$ and the red curve is a guide to the eye that marks the phase boundary in the presence of spin fluctuations. The dashed lime curve marks  $T_{hm}$, which is the phase boundary in the absence of spin fluctuations. The background color shows the Hall conductance calculated independently for a system of $L=W=50$, which conforms with $T_{sf}$.}
\label{Hall_PDiagram}
\end{figure}

In Fig.~\ref{Hall_PDiagram}, we draw a full phase diagram on the $m_0-T$ plane. Because the specific profiles of $\sigma$ and $S_0$ depend on the system size, the actual critical temperature $T_{sf}$ can be extracted by finite-size scaling. To this end, for a given set of variables, we calculate $\sigma$ as a function of $T$ for three different system sizes and identify the intersection of the three curves as $T_{sf}$ (see the inset of Fig.~\ref{Hall_PDiagram}). The critical temperature $T_{sf}$ ($T_{hm}$) calculated in the presence (absence) of spin fluctuations is depicted by red dots (dashed lime curve). We see that both $T_{sf}$ and $T_{hm}$ decreases monotonically with an increasing ratio of $m_0/\delta$. However, the discrepancy $\Delta T=T_{hm}-T_{sf}$, which measures the reduction of critical temperature due to spin fluctuations, reaches maximum around $m_0/\delta =0.75$; $\Delta T$ vanishes for both $m_0/\delta \rightarrow0$ and $m_0/\delta \rightarrow1$ limits.

Finally, we check the consistency of our conclusion by calculating the Hall conductance $\sigma_{xy}$ using the non-commutative Kubo formula with periodic boundary conditions, in which the Chern number is obtained directly from the real space rather than a momentum-space integral~\cite{Prodan2011Disordered,Prodan2012Quantum}. For a system of $L=W=50$, we numerically calculate $\sigma_{xy}$ and superimpose the result in Fig.~\ref{Hall_PDiagram}, where it exhibits a phase boundary that matches $T_{sf}$ remarkably well.

We stress that the mechanism of spin fluctuations studied in this Letter is entirely different from the ordinary magnon-electron scattering. First of all, we have considered the adiabatic regime such that spin fluctuations are frozen in time, whereas magnons are propagating spin waves. Second, spin fluctuations form a background random potential that scatters the electrons passively, while reversely, the excitation of spin fluctuations by electrons is ignored. Third, the physical picture of spin fluctuations persists up to $T_c$, whereas magnons are well defined only at low temperatures.

To close our discussion, we further remark that if adjacent magnetic layers are antiferromagnetically directed, the Dirac electrons will form an axion insulator rather than a QAH insulator below $T_c$, which has been realized in MnBi$_2$Te$_4$~\cite{Liu2020Robust}. Unlike the QAH insulators, the topological behavior in an axion insulator does not manifest in transport properties; instead it leads to quantized magneto-electrical responses~\cite{Qi2008Topological,Qi2009Inducing,Essin2009Magnetoelectric,Li2019Intrinsic}. However, by performing a similar analysis of spin fluctuations, we find that the coefficients of magneto-electrical responses only experience negligible changes.

In summary, we have demonstrated that spin fluctuations can play the role of a frozen random potential that leads to a significant reduction of the onset temperature of quantized transport in a magnetic TI. Even in the absence of structural disorders, considering the exchange gap at the mean-field level is insufficient to predict the critical temperature correctly. Our result provides an alternative explanation of the puzzling in recent experiments, and points out an unavoidable mechanism suppressing the quantized transport even in clean magnetic TIs.

We acknowledge insightful discussions with C. Z. Chen and Y. Z. You. This work was supported in part by the University of California, Riverside.
\hspace{3mm}

\bibliography{citation}

\begin{thebibliography}{36}
\bibitem [{dom()}]{domain}%
  \BibitemOpen
  \href@noop {} {}\bibinfo {note} {We have also considered the case of finite
  spatial correlation length, in which the lattice is divided into a
  superlattice consisting of $10\times10$ supercells. Magnetic spins are fully
  correlated, hence are uniform inside each supercell whereas spins from
  neighboring supercells are uncorrelated. We find no visible changes in our
  results.}\BibitemShut {Stop}%
\bibitem [{The()}]{Thermal}%
  \BibitemOpen
  \href@noop {} {}\bibinfo {note} {Typically, the Curie temperature of
  thin-film magnetic TIs is on the order of $2$meV, while the typical exchange
  field can be on the order of $100$meV.}\BibitemShut {Stop}%
\bibitem [{Edg()}]{EdgeWidth}%
  \BibitemOpen
  \href@noop {} {}\bibinfo {note} {Because the band gap in the QAH phase is
  $|\lambda|-|m_0|$, the width of an edge state is $w\approx
  v_f/(|\lambda|-|m_0|)$ with $v_f$ the Fermi velocity of the QAH edge states.
  When $m_0=0.8\delta$, $w$ is larger than the half-width of the sample, which
  partially overlaps with the opposite chiral edge state, resulting in an
  ill-quantized $\sigma$.}\BibitemShut {Stop}%
\end{thebibliography}

\end{document}